# Properties of MoO$_2$ and MoO$_3$ films prepared from the chemically driven isothermal close space vapor transport technique

O. de Melo,[1,2,3] Y. González,[2,4] A. Climent-Font,[5] P. Galán,[5] A. Ruediger,[4] M. Sánchez,[1] C. Calvo,[1] G. Santana,[3] V. Torres-Costa[2,5]

[1]Physics Faculty, University of Havana, 10400 La Habana, Cuba

[2]Departamento de Física Aplicada, Universidad Autónoma de Madrid. Cantoblanco 28049, Madrid, Spain

[3]Instituto de Investigación en Materiales, Universidad Nacional Autónoma de México, Cd. Universitaria, A.P. 70-360, Coyoacán 04510, México D. F.

[4]Institut National de la recherche scientifique, Centre Énergie, Matériaux, Télécommunications, 1650 Boulevard Lionel-Boulet, Varennes, Québec, J3X 1S2, Canada

[5]Centro de Microanálisis de Materiales. Universidad Autónoma de Madrid, 28049, Madrid, Spain

**Abstract.** Chemically – driven isothermal close space vapour transport was used to prepare pure MoO$_2$ films which were eventually converted to MoO$_3$ by annealing in air. According to temperature-dependent Raman measurements, the MoO$_2$/MoO$_3$ phase transformation was found to occur in the 225 – 350 °C range; no other phases were detected during the transition. A clear change in composition and Raman spectra, as well as noticeable modifications of the band gap and the absorption coefficient confirmed the conversion from MoO$_2$ to MoO$_3$. An extensive characterization of films of both pure phases was carried out. In particular, a procedure was developed to determine the dispersion relation of the refractive index of MoO$_2$ from the shift of the interference fringes the used SiO$_2$/Si substrate. The obtained refractive index was corrected taking into account the porosity calculated from elastic backscattering spectrometry. The Debye temperature and the residual resistivity were extracted from the electrical resistivity temperature dependence using the Bloch –

Grüneisen equation. MoO$_3$ converted samples presented very high resistivity and a typical semiconducting behaviour. They also showed intense and broad luminescence spectra, which were deconvoluted considering several contributions; and its behaviour with temperature was examined. Furthermore, surface photovoltage spectra were taken and the relation of these spectra with the photoluminescence is discussed.

1. Introduction

Molybdenum forms different oxides of formula MoO$_{3-x}$ with *x* ranging from 0 to 1. The most oxidized compound in the family, MoO$_3$, presents an orthorhombic structure, has n-type conductivity and a band gap of approximately 3.0 eV.[1] In the case of the MoO$_2$, the most reduced compound in the family, the stable phase presents a monoclinic structure. A metallic behavior, as predicted by theoretical calculations,[2] has been observed in MoO$_2$ although an optical band gap has been frequently found in its absorption spectra.[3] Among other important applications, both MoO$_2$ and MoO$_3$ present very interesting properties as catalysts, while oxygen deficient MoO$_x$ is being widely used as hole transport layer in either inorganic or organic solar cells. A comprehensive account of the applications of molybdenum oxides can be found elsewhere.[4]

Molybdenum oxides prepared with different techniques such as thermal evaporation,[5] sputtering[6,7] or pulsed laser deposition[8] for example, typically referred as MoO$_x$ in the literature, are composed by a mixture of different phases. This complicates the task of associating any measured property with a given phase, and has caused a large dispersion in the reported values for some properties of these oxides. In a previous paper, we have developed a method to obtain pure MoO$_2$ films by a chemically driven isothermal close space vapor transport technique (CD-ICSVT) using MoO$_3$ as the precursor source.[9] Moreover, the obtained MoO$_2$ films can be further oxidized to be converted into pure MoO$_3$. This provides an excellent opportunity for the study of the properties of these two pure phases. Now, in the present manuscript we present an extensive characterization of the electrical and optical properties of pure MoO$_2$ and MoO$_3$ samples which includes measurements of elastic backscattering spectrometry (EBS), electrical resistivity as a function of temperature, UV-VIS transmission and reflectance, surface photo-voltage (SPS), photoluminescence (PL) and Raman spectroscopies. Results are discussed and compared with those reported in the literature for samples grown with other techniques whenever this was possible.

2. Experimental methods

MoO$_2$ films were grown by the chemically driven isothermal close space vapor transport (CD-ICSVT) technique[9] in which a substrate is located at the same temperature and a few millimeter above a small amount of MoO$_3$ powder under a reductive gas flow at atmospheric pressure. Beyond its simplicity and compatibility with continuous processing, this technique has the advantage of producing pure MoO$_2$ samples instead of MoO$_x$ mixtures. MoO$_2$ films were grown at 570 °C onto SiO$_2$/Si or fused silica substrates under a flow of a H$_2$:Ar (1:5) gas mixture at a rate of 25 mL/min at atmospheric pressure. Some as-grown MoO$_2$ samples were converted into MoO$_3$ by oxidation in two different ways: i) using rapid thermal annealing at 400 °C for 1 min (similar annealings in N$_2$ gas did not lead to any change in the sample properties, while similar annealings in air at higher temperatures promoted, together with the phase transformation, a partial sublimation of the film) or ii) during Raman spectra acquisition as a function on temperature.

Raman spectroscopy was performed with a confocal optical microscope coupled to a modular Raman spectrometer from Horiba (iHR320). The light source was a continuous wave diode-pumped solid-state (CW-DPSS) laser with a wavelength of 473 nm (Cobolt Inc.), TEM 00, and linearly polarized, which was focused onto the sample with a 50 × microscope objective lens, NA 0.5; a power of 2.8 mW in front of the objective entrance was used. The Raman signal was dispersed into the spectrometer with a grating of 2400 lines/mm, then, a thermoelectrically cooled charge-coupled detector (CCD) (Synapse, Horiba Inc.) was used to collect the Raman spectra. The temperature was varied between 50 and 500 °C using a Linkam station THMS600 under ambient conditions.

To study the optical properties of the films, reflection and transmission spectra measurements in the 200–900 nm range were carried out in a Jasco V-560 UV–VIS double-beam spectrophotometer provided with an integrating sphere. PL measurements were carried out using the 325 nm wavelength He–Cd laser line (maximum output power of 16 mW) as excitation source. The sample emission was focused into an Acton SpectraPro 2500i spectrograph and detected by a photomultiplier tube. All the spectra were corrected taking into account the spectral response of the system. Surface photovoltage spectroscopy (SPS) was performed at room temperature in a lab-made automated experimental setup; comprising a halogen lamp coupled with a grating monochromator, a lock-in amplifier and an optical chopper.[10] The sample was located between a transparent indium–tin–oxide (ITO) electrode and a copper plate which acts as a sample holder and the bottom ground electrode. Measurements were carried on in soft contact mode configuration.[11]

Then, the surface photovoltage (SPV) signal measured as a function of incident photon energy provided the SPV spectra.

Elastic backscattered spectra (EBS) were obtained using alpha particles detected at a scattering angle of 170° with the alpha particles beam at three different energies of 2000, 3057, and 4268 keV in order to study the Rutherford scattering and also to promote resonant nuclear reactions with oxygen and carbon. The analysis of the EBS was performed using the computer code SIMNRA 6.06.[12] Values from non-Rutherford cross section for carbon and oxygen were generated using Sigma Calc 1.6 [13] while those for Si were provided by the file ASICH93A.RTR available in the SIMNRA computer code.[4]

High resolution scanning electron microscopy (HR-SEM) images were obtained by using an FE-SEM Hitachi S-4700 microscope. Resistivity measurements were carried out from 77 to 340 K in van der Pauw configuration.

3. Transformation from $MoO_2$ to $MoO_3$ by annealing in air

Raman spectra of an as-grown sample and a sample annealed at 400 °C for 1 min in air, onto fused silica substrates are shown in Fig. 1a. The most notorious modifications after the annealing are the emergence of the peaks at 820 and 1000 cm$^{-1}$, fingerprints of the $MoO_3$ structure,[14] and the disappearance of the peak at 740 cm$^{-1}$, characteristic of $MoO_2$. It reflects the phase change from $MoO_2$ to $MoO_3$ during the annealing process.

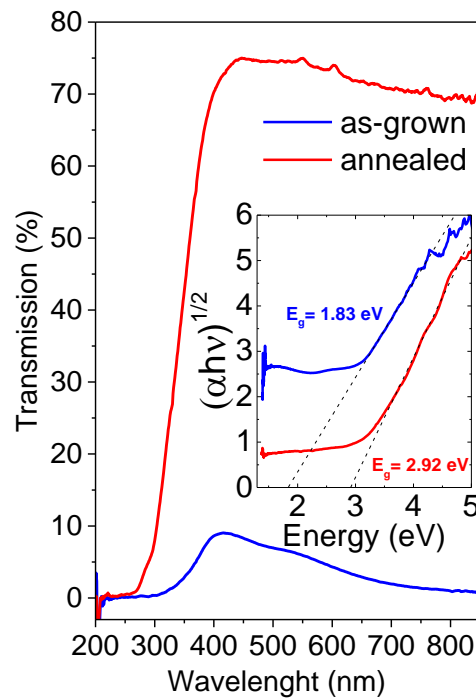

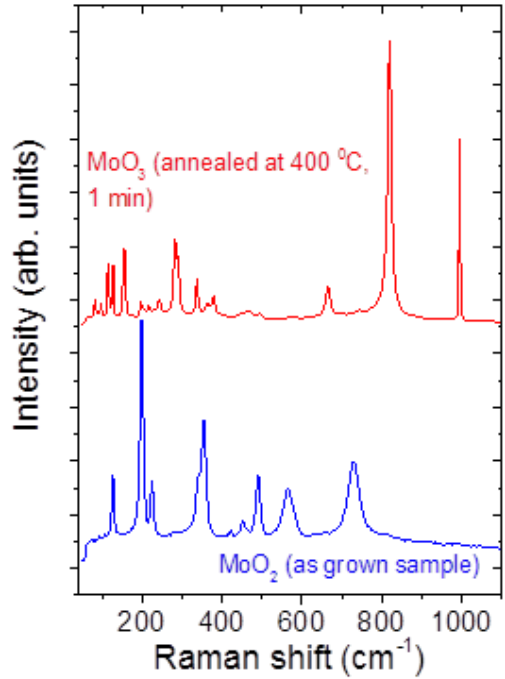

Fig. 1. a) Raman spectra of as-grown (blue curve) and annealed (red curve) samples. b) UV-VIS transmission spectra for an as-grown sample and for a sample subjected to a rapid thermal annealing at 400 °C for 1 min. In the inset, the Tauc plots, $(\alpha h\nu)^n$ $vs.$ $h\nu$ for both samples with $n = 1/2$ for both samples are displayed. The dotted lines in the inset of Fig. 1b indicate the extrapolated fundamental absorption to the respective optical band gap energies.

UV-VIS transmission spectra for the same samples are shown in Fig. 1b). A significant change in the transmission (that could be observed to the naked eye) as well as in the optical band gap, as consequences of the annealing, were observed. The $h\nu$ intercept of the linear region of the Tauc plot (in the inset), $(\alpha h\nu)^n$ $vs.$ $h\nu$ (with $n = 1/2$ corresponding to an indirect transition), indicated band gaps of 1.83 and 2.97 eV for the as-grown ($MoO_2$) and the annealed ($MoO_3$) samples, respectively. The result for $MoO_3$ is expected since a value near 3.0 eV has been typically reported in the literature[1,15] for the band gap of this material. With respect to the $MoO_2$ sample, it is worth to comment that a clear absorption edge should not be expected since a metallic behavior has been predicted from theoretical band structure calculation. However, besides our present result, several authors have reported the existence of a band gap for reduced molybdenum oxides. For example, band gap values of 2.4 and 2.7 eV have been reported in reduced $MoO_3$ films although those samples really contained a mix of different $MoO_x$ (2 < x < 3) phases.[5] In another report,[3] a band gap of 4.22 eV was found for MoO$_2$ but it cannot be compared with our value since it was determined using $n = 2$ in the Tauc plot (our spectra return a band gap of around 3.5 eV if *n* is taken as 2). Our result confirms an optical band gap existence for pure MoO$_2$. The existence of an optical band gap in a material in which the Fermi level crosses electronic bands, as in the present case, is puzzling and deserve more experimental and theoretical and experimental investigations. A decrease in the transmission is observed for longer wavelengths in MoO$_2$, probably due to free electron absorption, which seems to confirm a mixed metallic/semiconductor structure character of this material.

To study the evolution of the phase transition, Raman spectra were measured as a function of temperature. As it can be noticed in Fig. 2, the phase transition is observed in the range between 225 °C (the 820 cm$^{-1}$ peak emerges) and 350 °C (740 cm$^{-1}$ peak disappears). A detailed analysis of these spectra shows that no other phase was present at any temperature, not even during the phase transition. Other authors have reported[16] the phase transition taking place between 230 and 490 °C for MoO$_2$ nanosheets. As it can be noted, the lower temperature limit of the transition in that paper is similar to the one reported here. However, they claim not having observed a complete conversion since phase coexistence persisted up to 490 °C, the highest measured temperature in their experiment. Probably, it is due to the consideration of the peak at near 340 cm$^{-1}$ as belonging to MoO$_2$ phase while there is a peak very close to it that actually belong to MoO$_3$.[11] The spectrum at the top was measured after cooling the sample, it can be noted that the MoO$_3$ phase is preserved, indicating the expected not reversible character of the MoO$_2$/MoO$_3$ phase transformation in air.

To determine the composition profiles of the samples, EBS were taken with three different incident alpha particles energies. At 2000 keV all elements of interest follow Rutherford scattering cross sections and there was a good depth resolution (better than 4 x 10$^{16}$ atom/cm$^2$, about 8 nm) in the layer. The EBS spectra for MoO$_2$ and MoO$_3$ samples and the corresponding depth profiles are shown in Fig. 3 a- d). As expected, the stoichiometry of the as-grown sample corresponds with that of the MoO$_2$. By using incident alpha particles with energy of 4268 keV, at which the elastic resonance 12C $^{12}$C($\alpha$, $\alpha_0$)$^{12}$C, occurring at 4258 keV provides a good sensitivity for the detection of carbon, a surface carbon contamination was determined to be equivalent to a thickness of 7 x 10$^{15}$ atom/cm$^2$, probably originating in the graphite boat used for the growth. (The spectra taken with incident energies other than 2000 keV are shown in the supplementary material). Moreover, EBS results indicate that the MoO$_2$ layer presents some roughness. This lack of thickness uniformity is manifested in the EBS spectra as a relatively wide interface with the substrate, as observed in the deduced depth profiles, although this interface may be quite sharp in reality. Carbon was also detected on the surface of the annealed sample. In this case, the annealing in air of the $MoO_2$ layer results in the incorporation of oxygen in the film. This is evident in the depth profiles of Fig. 3 c) and d) in which the atomic concentration of Mo change from 0.66 (Mo/O: 1:2) to 0,75 (Mo/O: 1/3) approximately.

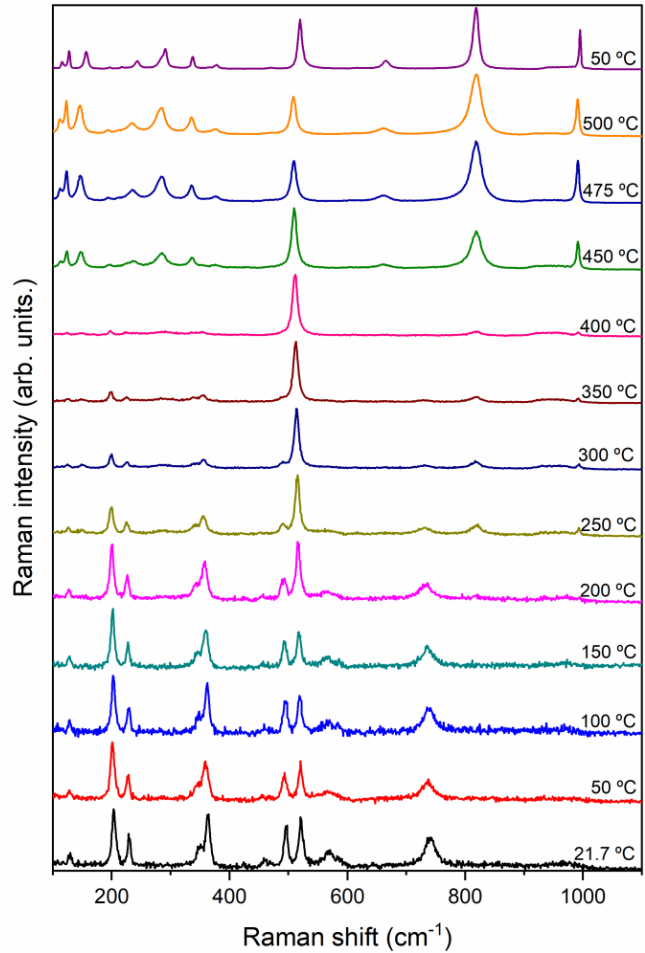

Fig. 2. Evolution of Raman spectra as a function of temperature. The transition from $MoO_2$ to $MoO_3$ is observed in the range between 225 - 350 °C. The spectrum at the top was measured after cooling the sample. The normalized spectra are shifted in vertical for illustration purpose only

Other conclusion to be drawn from the EBS spectra is that the $MoO_2$ samples were porous. In fact, the thickness calculated considering the amount of Mo in the layer and the atoms density of $MoO_2$ is much smaller than the actual (measured) thickness. This means that the atoms density of the films is smaller than that of bulk $MoO_2$. With both values, the film thickness calculated from the EBS spectrum and that measured with a profilometer, the porosity was evaluated (see supplementary material) to be around 0.5; this value was used to correct the measured refractive index of $MoO_2$ as can be seen in the next paragraph.

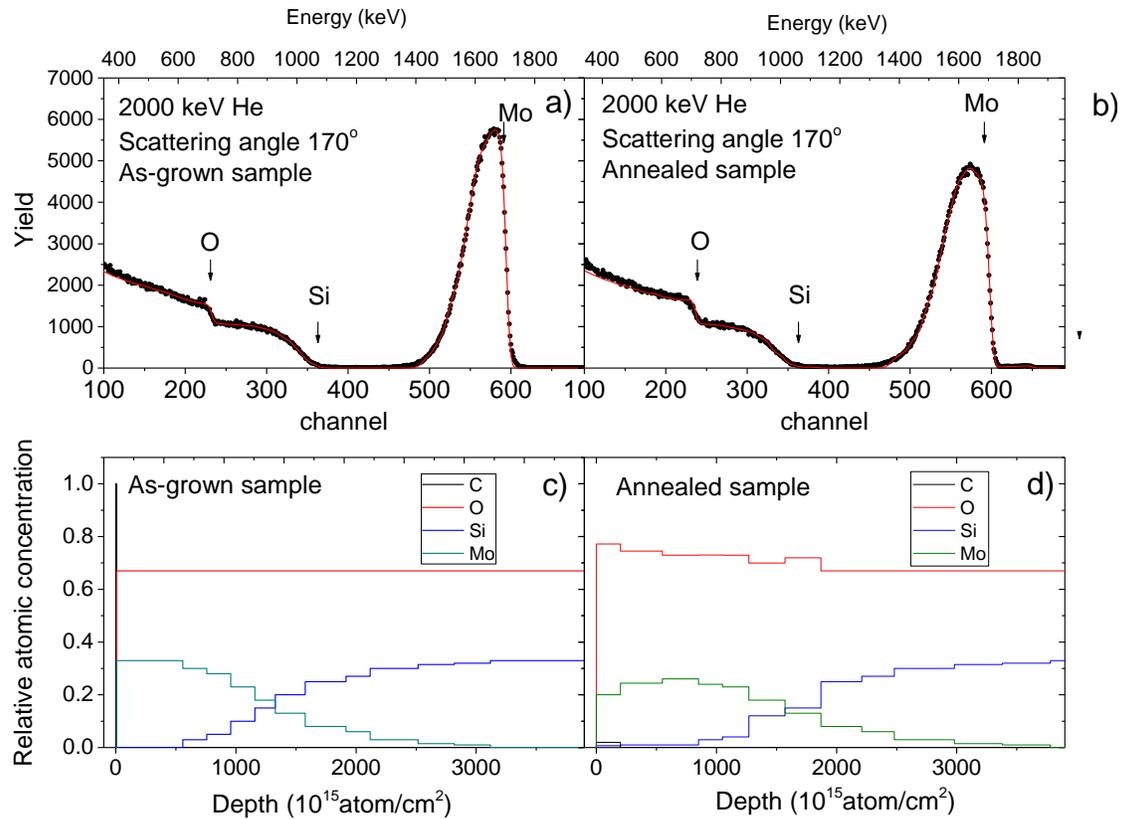

**Fig. 3.** Rutherford backscattering spectra for an as-grown sample (a) and an annealed one (b) two samples grown at the same conditions as those in Fig. 5

4. The refractive index of $MoO_2$

Reflectance spectroscopy in the UV-VIS region is a useful tool for determining the optical properties and thickness of thin films. Even when the absolute value of the reflectance is hard to be measured (i.e., because of sample size), fitting the position and relative amplitude of interference fringes allows the determination of the refractive index of films of known thickness. In general, interference fringes only arise for relatively transparent films in which the radiation entering at the external surface of the films, can reflect at the internal interface and return to the external surface without significant loss of intensity. Due to the absorbing character of $MoO_2$, only very thin films present appreciable transmission, but at the same time, in very thin films the appearance of interference fringes is excluded because the small optical path difference introduced by the film. We have devised the following approach to measure the refractive index of $MoO_2$. Very thin $MoO_2$ films, in the range of tens of nanometers were grown onto $SiO_2$/Si substrates, which present their own interference fringes due to the relatively large thickness and transparence of the $SiO_2$ film. The very thin $MoO_2$ films grown on top of these substrates do not show interference fringes of their own, but they induce a noticeable shift of the substrate's fringes. This effect is illustrated in Fig. 4a in which the reflectance spectra of the $SiO_2$/Si substrate and that of the grown $MoO_2$/$SiO_2$/Si structure are shown. Then, simulating the reflectance spectra of thin $MoO_2$ films on top of the $SiO_2$/Si substrate by the transfer matrix procedure considering two-layer interference, and fitting the induced spectral shift for a sample of known thickness, the refractive index can be determined. Fig. 4b shows the effect of the increasing $MoO_2$ layer thickness on the simulated reflectance spectra of a $SiO_2$(305nm)/Si substrate. The interference fringes are clearly shifted even for $MoO_2$ layers less than 2 nanometers thick. The inset in Figure 4b details the evolution of the interference maximum initially in the pristine $SiO_2$ (305nm)/Si substrate at 435nm as a function of $MoO_2$ thickness.

In order to determine the (complex) refractive index of the $MoO_2$ layers and its dispersion relation, using the above procedure, the reflectance spectrum of a $MoO_2$/$SiO_2$(305nm)/Si structure of known $MoO_2$ thickness was simulated and fitted to the experimental measurement. The actual $MoO_2$ layer thickness was determined by cross-sectional SEM (Fig. 5a), the well-known values of $SiO_2$ and Si refractive indexes were introduced into the simulation as well. Fig. 5b shown the experimental and simulated reflectance spectra of the 100 nm $MoO_2$ layer grown on a $SiO_2$(305nm)/Si substrate, while in Fig. 5c the real and imaginary parts of the refractive index of $MoO_2$ obtained from this fitting procedure are showed. Again, the goal of this simulation was to properly fit the position and height of the interference fringes, not their absolute intensity value.

A remarkably low refractive index at the red edge of the spectrum is observed in Fig. 5c. Taking into account that the refractive index obtained by this method correspond to a effective refractive index of the $MoO_2$ film, such a low value suggests a low $MoO_2$ density due to the porosity of the layer, in accordance with the EBS measurements commented above. The steep increase in the extinction coefficient and the anomalous dispersion of the refractive index below 400 nm suggest the presence of an absorption edge as observed in transmission measurements above.

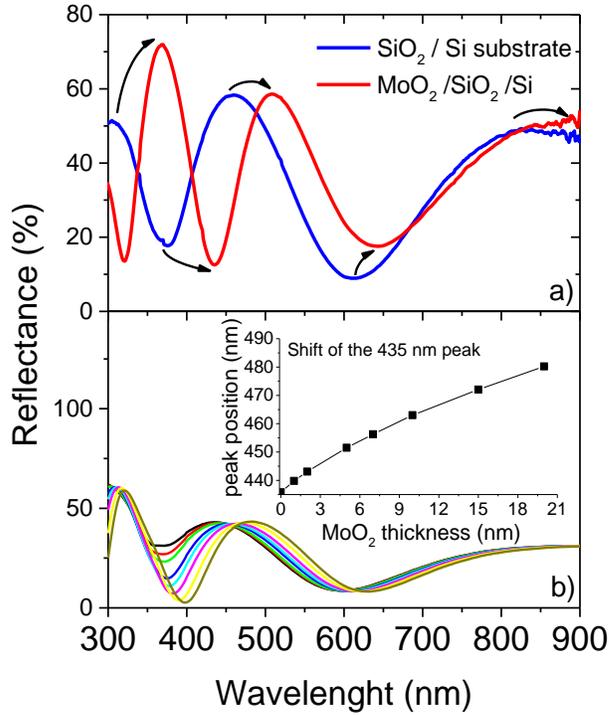

Fig. 4. Reflectance spectra for MoO₂/SiO₂(305nm)/Si structures. A clear shift of the interference fringes is observed even for MoO₂ layers less than 2 nm thick. In the inset: evolution with MoO₂ thickness of the interference maximum initially at 435 nm.

The effective refractive index of porous MoO₂, $n_{MoO_2}^{eff}$, can be approximated, using effective medium theory,[17] as a combination of the refractive indices of the MoO₂, $n_{MoO_2}$, and air, $n_{air}$:

$$n_{MoO_2}^{eff} = n_{air}p + n_{MoO_2}(1-p)$$

Here, $p$ represents the porosity, i.e. the void fraction of the total volume. From this equation, and using the porosity determined by the EBS spectra, the refractive index of MoO₂ was calculated. The corrected refractive index is shown in Fig. 6.

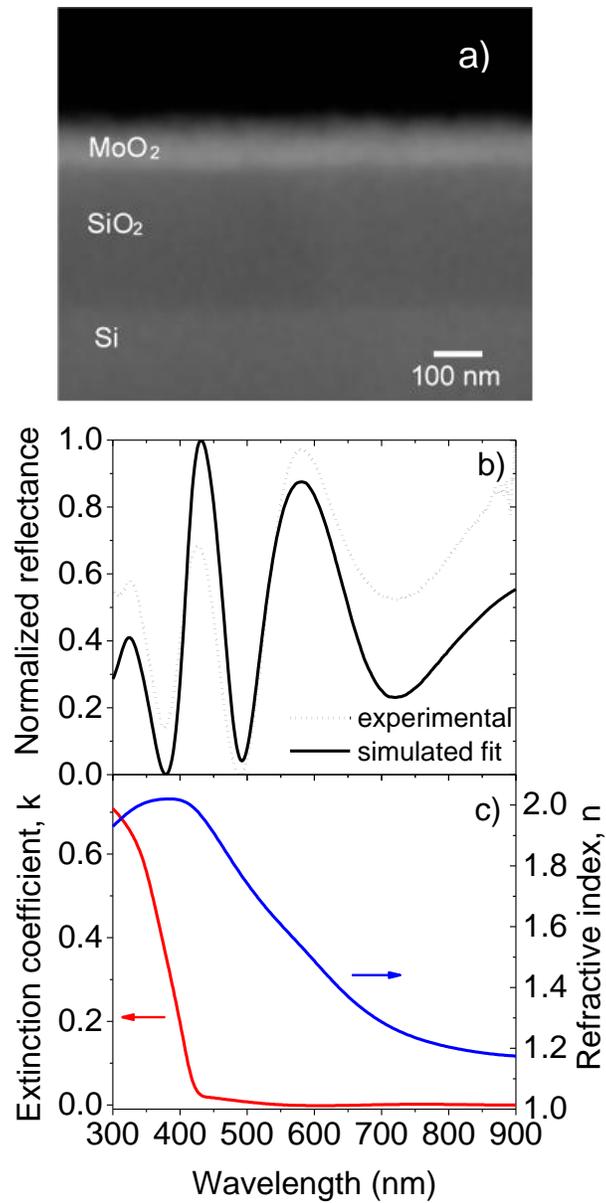

Fig. 5. a) Cross section SEM image of the film used for determining the refractive index of $MoO_2$. b) Experimental and simulated reflectance spectra of a 100 nm $MoO_2$ layer grown on the $SiO_2$(305nm)/Si substrate and c) the refractive index of $MoO_2$ obtained from the fitting procedure

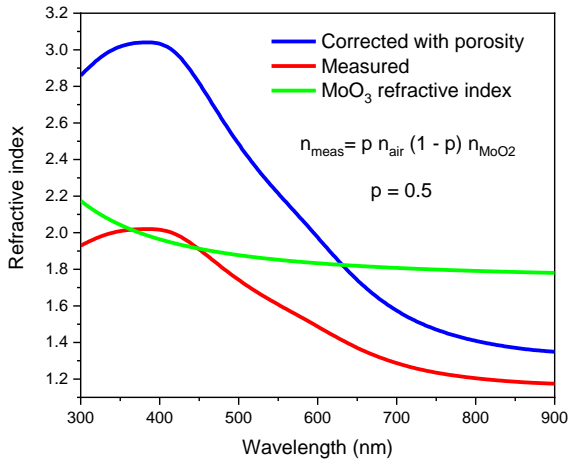

Fig. 6. Refractive index of $MoO_2$. Measured (effective) and corrected with porosity are shown. The refractive index of $MoO_3$ is also reported for comparison.

In the only previous report found in the literature, the refractive index of electro-deposited $MoO_2$[18] was measured at a wavelength of 632.8 nm. The authors determined values ranging between 1.7 and 2, depending on the morphology and thickness of the films. Our measured dispersion relation provides a value of 1.81 at this particular wavelength which is inside that range. In Fig. 6., for comparison, the refractive index dispersion relation of $MoO_3$ is also presented.

The determined dispersion relation of the effective refractive index of $MoO_2$ allowed the estimation of the layer thickness of the films from simple reflectance measurements (using the inverse process than that described above for determining the refractive index from a sample with known thickness). For two relatively thick samples with thickness of 65 and 120 nm (measured by profilometry) this method yields values, of 50 and 100 nm, respectively, which represents a relatively good agreement between both methods.

5. Temperature dependence of the electrical parameters

5.1 Electrical resistivity and Bloch–Grüneisen parameters of $MoO_2$

A very low resistivity in the $10^{-4}$ Ohm-cm range was typically determined by the four probe method for the as-grown $MoO_2$ samples. The dependence of the resistivity with temperature from 77 K to RT is shown in Fig. 7 in which a typical metallic behavior with an approximately linear increase of resistivity at high temperatures was observed. The temperature dependence of the resistivity for metals is commonly described by the Bloch–Grüneisen (BG) integral equation:

$$\rho(T) = \rho_0 + K \, \frac{T^n}{T_D^{n+1}} \int_0^{\frac{T_D}{T}} \frac{x^n}{(e^x - 1)(1 - e^x)} dx$$

in which, $\rho_0$ is the residual low temperature resistivity, K is a constant proportional to the electron-phonon coupling constant, $T_D$ is the Debye temperature and the exponent *n* can take the values 2, 3 or 5 depending on the electron dispersion mechanism. The $\rho - T$ curve was fitted quite well using the above equation and C and $T_D$ as fitting parameters. $\rho_0$ was taken as the low temperature resistivity and different values of n were tried. As a result of the fitting procedures the following values were obtained: $\rho_0 = 4.4 \times 10^{-5} \, \Omega - cm$, $T_D = (617 \pm 6) \, K$, $C = (8.5 \pm 1.5) \times 10^{-4} \, \Omega - cm - K$ and n = 5. Attempts to use n = 2 or 3 returned unrealistic high values of $T_D$. This is a very clear metallic behavior with the electron dispersion limited by the electron-phonon interaction. The value of the high temperature limit of the slope of the $\rho - T$ curve,[19] $\beta = C/4T_d^2$, was calculated as 0.34 $\mu\Omega - cm/K$. In ref. 20 the source-drain resistance of a $MoO_2$-$MoS_2$ heterostructures was found to increase with temperature in a way similar to the BG behavior; they found $T_D = 709.3 \, K$ and n = 2. But the resistivity was not measured and the material was not pure $MoO_2$. Other authors have reported the metallic behavior of $MoO_2$[21] but, in the best of our knowledge, our report is the first one of the Bloch–Grüneisen parameters of $MoO_2$. For comparison, values of $\rho_0$, $T_D$ and $\beta$ reported in the literature for different materials are shown in Table I.

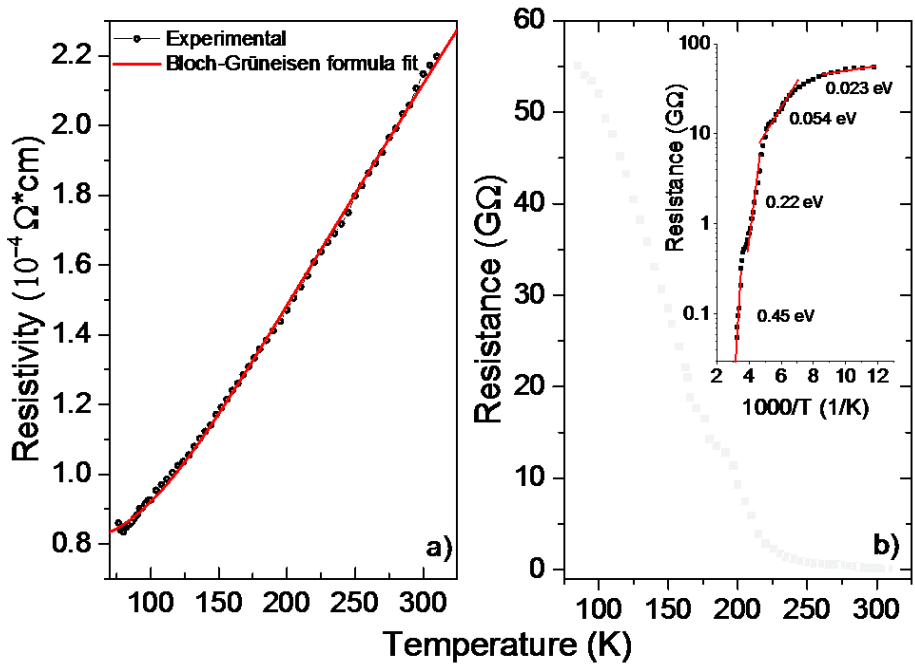

Fig 7. a) Resistivity as a function of temperature for a MoO$_2$ sample showing the typical metallic behavior that fit quite well the Bloch- Grüneisen equation and b) resistance as a function of temperature of a MoO$_3$ sample. In the inset, the Arrhenius plot of the resistance shows activation energies on different temperature ranges.

Table I. Values of residual resistivity ($\rho_0$), Debye temperature ($T_D$) and temperature coefficient ($\beta$) reported in the literature for different materials

|  | $\rho_0\ (\Omega-cm)$ | $T_D\ (K)$ | $\beta\ (\mu\Omega-\frac{cm}{K})$ | Ref. |
|---|---|---|---|---|
| MoO$_2$ | 8.06 x 10$^{-5}$ | 617 | 0.34 | This work |
| Cr | (3.4-381)x10$^{-6}$ | 384-464 | 0.041-0.14 | [19] |
| ITO | (8.53-11.7)x 10$^{-4}$ | 1018-1085 | 0.6-1.0 | [22] |
| In$_2$O$_3$ |  | 700 |  | [23] |
| ZnO:Al | (3.2-7.9) x 10$^{-4}$ | 1075- 1240 | 0.76- 1.1 | [24] |
| Silver | 1.59 x10$^{-6}$ | 221 | 0.0038 | [25,26] |
| Copper | 1.68 x10$^{-6}$ | 310 | 0.0039 | [16] |

As it can be observed in Table I, the resistivity of MoO$_2$ is almost two order of magnitude larger than common metals like silver or copper but is smaller than transparent conductive oxides like ITO or ZnO:Al. The temperature coefficient and the Debye temperature are higher for MoO$_2$ with respect to the metals but is smaller than those of ITO, ZnO:Al and In$_2$O$_3$. These transparent conductive oxides present also a metallic behavior but they have a clear band gap in the band structure.

5.2. Temperature dependence of the resistance in MoO$_3$

In contrast to MoO$_2$, the samples converted to MoO$_3$ were found to be highly resistive. This made it impossible to obtain reliable values of the resistivity using a four probes measurement as in the Van der Pauw technique; a two probe one was used instead. Although in this kind of measurement, the contact resistance can induce overestimated values for the resistivity of the samples, the qualitative behavior of the sample resistance can be analyzed. Fig. 7b show the temperature dependence of the resistance of a MoO$_3$ film. A decrease of the resistance with increased temperature is observed in all the temperature range as an indication of the semiconductor material behavior. In the inset, the Arrhenius plot shows the presence of four linear regions with activation energies of 0.45, 0.22 and 0.054 and 0.023 eV, respectively. They could be associated with the ionization of relatively deep levels into the band gap, presumably originated by oxygen vacancies. The presence of deep energy levels is also reflected in the photoluminescence properties as can be described below.

5   Photoluminescence and surface photovoltage of MoO$_2$ samples converted to MoO$_3$

While MoO$_2$ films did not show appreciable PL at any temperature, a strong and broad emission was observed in samples annealed to MoO$_3$. Typical PL spectra at 12 K and RT are shown in Fig. 8a, b) (all the spectra can be accessed at the supplementary information). The spectrum was fitted with six Gaussian contributions located at 400.9 (3.09), 410.6 (3.02), 427.2 (2.90), 463.4 (2.68), 527.7 (2.35), and 618.0 (2.0) nm (eV) in the 12 K spectrum. Several MoO$_3$ samples were measured and their spectra were fitted with relatively similar contributions. Other authors[27] ascribe the blue emissions ranging from 400 to 470 nm to near band gap emissions, and associate the rest of the emission bands to defects states in the band gap. In an effort to elucidate the origin of the different emission bands we studied the evolution of the PL with temperature. The behaviour of the integrated area and the energy position of the PL peaks are shown in Fig. 8 c and d.

With increasing temperature, the intensity of the peaks remains relatively constant up to around 140 K, then it starts to decrease exponentially (except the peak at 527.7 nm whose intensity increase and the peak at 618.0 nm that remains practically unchanged). This temperature dependence of the peaks intensity causes an overall green shift of the RT luminescence as shown in Fig. 8b. The activation energy of the luminescence quenching is of around some tens of meV (27, 42.4, 52, 21.3 meV for peaks P1. P2, P3 and P4, respectively with P3 and P4 having activation energies close to those observed for thermally induced conductivity). The wavelength position of the peaks, even those in the blue region of the spectrum, is practically constant with temperature. Since near band gap PL peaks positions are expected to decrease with increasing temperature following the band gap behaviour, the above result indicate that all the observed emissions are originated by defects band gap states, in contrast with the assumption in ref.[25].

To study the influence of the band gap defect states in the electrical properties, SPV spectrum was measured at RT in the range 350 – 600 nm. SPS technique relies on the measurement of the change in the surface potential as a result of the transfer or redistribution of charges due to the external illumination. The absorbed photons induce the formation of free carriers via band-to-band transitions and/or release captured carriers via trap-to-band transitions. A typical SPV spectrum is displayed (black line) in Fig. 8 b. As can be observed, a surface photovoltage response exists in the all PL emission range. The later confirm that PL origin is originated in band gap defect states.

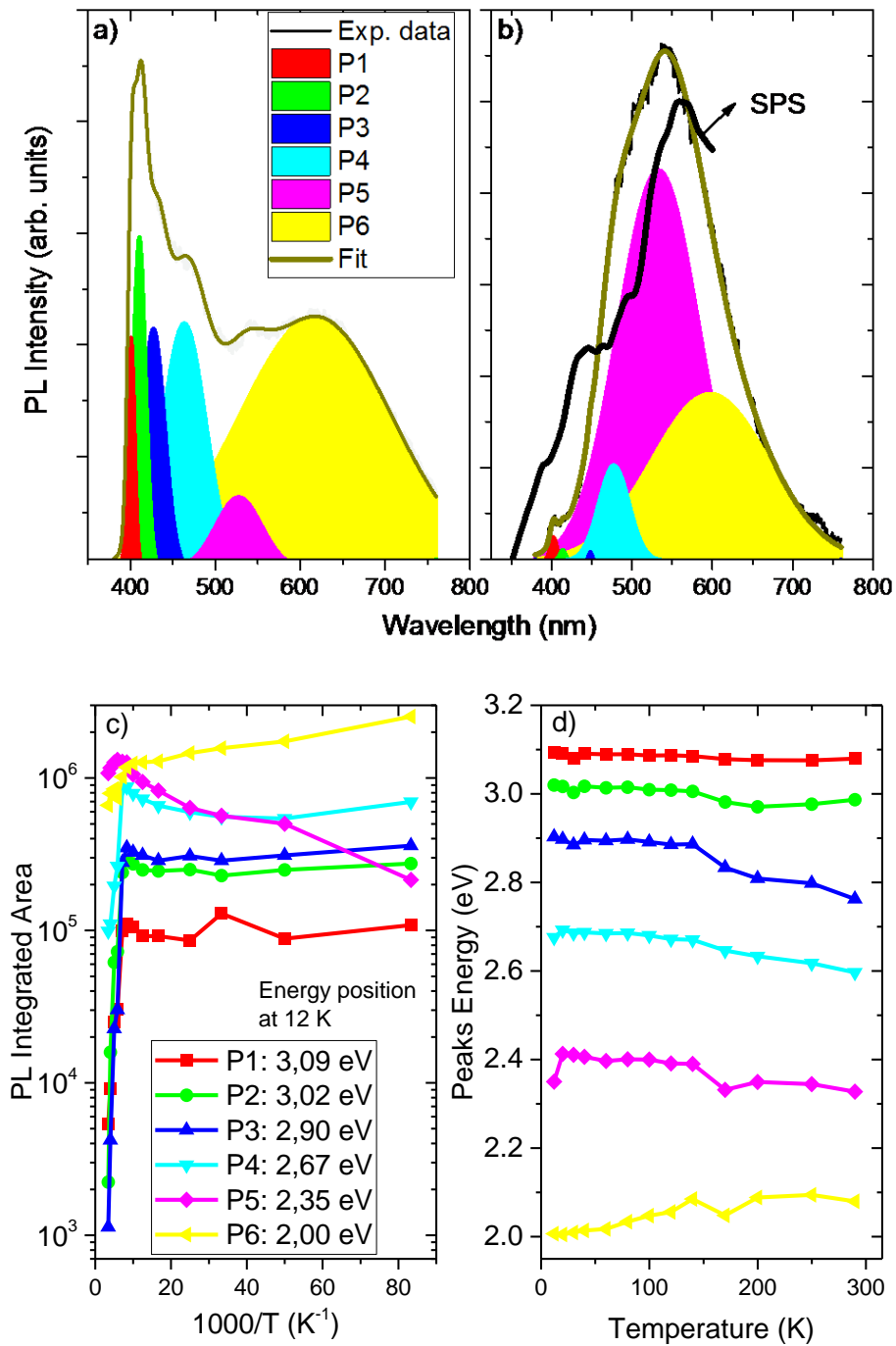

Fig. 8. Typical PL spectra at 12 K (a) and RT (b) including Gaussian contribution used to fit the experimental data (the SPV spectrum by a black line is also showed in b. Temperature behaviour of the integrated area of the different contributions(c) and the peak energy position (d)

Conclusions

Pure $MoO_2$ and $MoO_3$ films have been obtained; $MoO_2$ by using chemically driven close space vapor transport and $MoO_3$ by annealing in air the as-grown $MoO_2$ films. The $MoO_2$/$MoO_3$ phase transition was found to occur in the 225 - 350 °C range. We reported an optical band gap for pure $MoO_2$ of 1.83 eV assuming an indirect transition; this value seems to be more realistic than others reported in the literature, due to the absence of mixed phases in our material. After the simulation of substrate interference fringes shift resulting from the deposition of $MoO_2$, the dispersion relation of the refractive index of $MoO_2$ is reported in this paper by the first time. The measured value was corrected considering the porosity estimated by Rutherford backscattering spectra.

Electrical resistivity measurements as a function of temperature confirmed the metallic character of $MoO_2$ and allowed to fit the Bloch–Grüneisen equation to extract the relevant parameters as the Debye temperature, the residual resistivity, and the high temperature limit of the $\rho - T$ curve slope. Although a metallic behavior had previously been observed in MoO2, our fit of the Bloch –Grüneisen formula allowed to determine the Debye temperature of the material. The photoluminescence of the samples converted to $MoO_3$ was measured as a function of temperature. A broad wide spectra producing white luminescence was observed at low temperature and fitted with six Gaussian contributions. With increasing temperature, the higher energy peaks partially quenched and the color of the overall luminescence changed to green. The energy position of the peaks, comprising those with energies in the blue region, near the band gap, remain practically constant. This indicates that all the observed peaks are originating from defect states in the band gap.

Acknowledgements. OdM and YG gratefully acknowledge the support from the program ''Cátedras de Excelencia de la Comunidad de Madrid (2016-T3/IND-1428). OdM. and GS thanks the support of UNAM/DGAPA/PREI program 2018. A.R. gratefully acknowledges financial support from NSERC through a discovery grant.